\documentclass[aps,preprint,showpacs,showkeys]{revtex4}
\usepackage[pdftex]{graphicx}
\usepackage{amssymb}
\usepackage{bm}
\begin{document}
\setcounter{page}{1}
\title[]{The Lifetime of the Universe}
\author{Don N. \surname{Page}}
\email{don@phys.ualberta.ca}
\thanks{Alberta-Thy-14-05, MIFP-05-25, submitted to the
Journal of the Korean Physical Society for the proceedings of
the 9th Italian-Korean Symposium on Relativistic Astrophysics,
Seoul, South Korea, and Mt. Kumgang, North Korea, 2005 July 19-24.}
\affiliation{Department of Physics, University of Alberta,
Edmonton, Alberta T6G 2J1, Canada; \\
George P. and Cynthia W. Mitchell Institute for Fundamental Physics,
Texas A \& M University, College Station, Texas 77843-4242, USA; \\ and \\
Asia Pacific Center for Theoretical Physics (APCTP), Hogil Kim Memorial Building
\#519, POSTECH, San 31, Hyoja-dong, Namgu, Pohang, Gyeongbuk 790-784, Korea}
\date[]{2005 September 30}

\begin{abstract}

Current observations of the fraction of dark energy and a lower limit on its
tension, coupled with an assumption of the non-convexity of the dark energy
potential, are used to derive a lower limit of 26 billion years for the future
age of the universe.  Conversely, our ordered observations, coupled with an
assumption that observers are smaller than the universe, are used to argue for
an upper limit of about $e^{10^{50}}$ years if the universe eventually
undergoes power-law expansion, and an upper limit of only about $10^{60}$ years
left for our universe if it continues to expand exponentially at the current
rate.

\end{abstract}



\maketitle

\section{Introduction}

A long-standing question in cosmology is the future lifetime of the universe.
With traditional Friedman-Robertson-Walker dust models, it was thought that the
universe would last forever if the deceleration parameter were $q_0 \leq 1/2$
and would recollapse in finite time if $q_0 > 1/2$.  However, these dust models
have been consigned to the dustbin, and the question has become
quintessentially more difficult.

Here I shall give simple arguments for getting both lower and upper limits on
the future lifetime of the universe.

\section{Lower limits on the lifetime of the universe}

The recent evidence for cosmic acceleration \cite {Ries,Perl} is often
interpreted as that for a positive cosmological constant and eternal expansion
of the universe (e.g., in early WMAP press releases \cite{WebWMAP}).

However, many people have emphasized that the dark energy giving the current
acceleration could be a scalar field $\phi$ with a potential $V(\phi)$ that
could go negative and lead to a collapsing universe.  See, for example,
\cite{Star,KL,ASS,KKLLS,WKLS,Per}.  Here I shall summarize a simplified form of
the argument and get a $w$-dependent lower limit on the lifetime if $V(\phi)$
is not convex.

Scholars have dusted off the dust model by adding a scalar field $\phi(t)$ with
potential $V(\phi)$.  With $8\pi G/3 = c = 1$, $\dot{f} \equiv df/dt$,
$f' \equiv df/d\phi$, $\dot{\phi} \equiv -v$, and $K \equiv {1\over 2}\dot{\phi}^2
\equiv {1\over 2} v^2$,
a spatially-flat ($k=0$) FRW scalar-dust universe obeys
\begin{eqnarray}
&&H^2 \equiv \left({\dot{a}\over a}\right)^2 = \rho_m + \rho_s
    = {\mathrm{const.}\over a^3} + {1\over 2}\dot{\phi}^2 + V(\phi)
    = \rho_m + K + V, \nonumber \\
&&\ddot{\phi} + 3H\dot{\phi} + V' = 0,
\label{eq:1}
\end{eqnarray}
where $\rho_m = \mathrm{const.}/a^3$ is the energy density of the dust,
and $\rho_s = {1\over 2}\dot{\phi}^2 + V(\phi)$ is the energy density
of the scalar field, which has pressure $P_s = {1\over 2}\dot{\phi}^2 - V(\phi)$.

These equations imply
\begin{eqnarray}
&&\dot{\rho}_m = -3H\rho_m, \nonumber \\
&&\dot{\rho}_s = -3H(\rho_s + P_s) = -3H(2K),
\label{eq:2}
\end{eqnarray}
which further imply
\begin{eqnarray}
\dot{H} = -\:{3\over 2}(\rho + P) = -\:{3\over 2}(\rho_m + 2K)
        = -\:{3\over 2}(H^2 + K - V) < 0.
\label{eq:3}
\end{eqnarray}

Also, $\dot{v} \equiv -\ddot{\phi} = V' - 3Hv$, so if $V'$ stays larger
than $3Hv$ as $\phi$ decreases, $v$ and $K \equiv {1\over 2}v^2$ will continue
to increase, keeping $-\dot{H}$ bounded away from zero.  Then $H$ will
decrease indefinitely to $-\infty$, and the universe will end in a Big Crunch.

Eliminate the dust density by using $\dot{\rho}_m = -3H\rho_m$.
Then one has three coupled first-order ordinary differential equations,
$\dot{\phi} = -v$, $\dot{v} = V' - 3Hv$, and
$\dot{H} = -{3\over 2}(H^2 + {1\over 2}v^2 -V)$.  The time does not appear explicitly,
so one can eliminate it and use $f' \equiv df/d\phi = \dot{f}/\dot{\phi}
= - \dot{f}/v$ to get two coupled first-order ordinary differential equations:
\begin{eqnarray}
&&v' = 3H - {V'\over v}, \nonumber \\
&&H' = {3(H^2-V)\over 2v} + {3\over 4} v.
\label{eq:4}
\end{eqnarray}
Or, we can use the first of these two equations to eliminate $H$ and get
one second-order ordinary differential equation, simpler when written in terms
of $K \equiv {1\over 2}v^2$ rather than $v$:
\begin{eqnarray}
K'' = {(K'+V')(3K'+V')\over 4K} + {9\over 2} K - 9V - V''.
\label{eq:5}
\end{eqnarray}

In principle, we need two boundary conditions, e.g., $v$ and $H$ (or $K$ and $K'$)
at some $\phi$, or $\phi$ and $K$ when the dimensionless density parameter $\Omega_m$
reaches its present value $\Omega_{m0}$, $\Omega_m = (H^2-K-V)/H^2 = \Omega_{m0}$.
However, generic data at $\Omega_m = \Omega_{m0}$ will evolve back to $K=\infty$
at $H=\infty$ (at the Big Bang, $a=0$).  Boundary conditions arising out of
early-universe inflation would suggest restricting to finite $K$ at $H = +\infty$,
which leaves only one arbitrary parameter, say the initial scalar field value,
$\phi = \phi_i$, at the beginning of the universe, at $H=+\infty$.

If $V(\phi) \approx V_i - V'_i(\phi_i-\phi) + {1\over 2}V''_i(\phi_i-\phi)^2$
for $0 < \phi_i-\phi \ll 1$, then Eq. (\ref{eq:5}) is solved with
\begin{eqnarray}
K \approx {1\over 3}V'_i(\phi_i-\phi) - {3\over 10}(6V_i+V''_i)(\phi_i-\phi)^2
\label{eq:6}
\end{eqnarray}
for $0 < \phi_i-\phi \ll 1$.  This gives
\begin{eqnarray}
t = \int_{\phi_i}^{\phi}{-d\tilde{\phi}\over\sqrt{2K(\tilde{\phi})}}
 \approx \sqrt{{6(\phi_i-\phi)\over V'_i}},
\label{eq:7}
\end{eqnarray}
implying $\phi \approx \phi_i - {1\over 6} V'_i t^2$ and
\begin{eqnarray}
H = {K'+V'\over\sqrt{18K}}\ \approx \ \sqrt{2V'_i\over 27(\phi_i-\phi)}
 \ \approx \ {2\over 3t},
\label{eq:7b}
\end{eqnarray}
which describes the initial dust-dominated phase.

Then for a chosen initial value $\phi_i$, evolve $\phi_i-\phi$ to the point where
$\Omega_m = 1 - (K+V)/H^2 = 1 - 18K(K+V)/(K'+V')^2 = \Omega_{m0}$,
and there evaluate
\begin{eqnarray}
w \equiv {P_s\over \rho_s} = {K-V\over K+V} = w(\phi_i,\Omega_{m0}).
\label{eq:8}
\end{eqnarray}

Thus for a given scalar field potential function $V(\phi)$ and given present values
$\Omega_{m0}$ and $w_0$, solving $w(\phi_i,\Omega_{m0}) = w_0$ for $\phi_i$ gives
$\phi_i = \phi_i(\Omega_{m0},w_0)$ and hence a unique model.  Then one can solve
for the ratio of the future time until the Big Crunch
to the past time since the Big Bang,
\begin{eqnarray}
R = R(\Omega_{m0},w_0) = {t_c\over t_0}
  = {\mathrm{future\ time\ to\ crunch}\over\mathrm{past\ time\ since\ bang}}.
\label{eq:9}
\end{eqnarray}

If $V(\phi)$ is convex with a nearby cliff, in principle the universe could fall
off the cliff and recollapse in the next minute.  However, with the hope that you
will be able to read the rest of this paper, I shall assume that this is implausible.

Therefore, for concreteness, assume that $V(\phi)$ is not convex.  Then the smallest
$R(\Omega_{m0},w_0) = t_c/t_0$ would be for the case of a linear potential,
$V(\phi) = \beta\phi$.

With $8\pi G/3 = c = 1$, the dimensions of mass, length and time are the same,
the scalar field $\phi$ is dimensionless, and $\beta$ has the dimensions of $V$,
which is that of energy per length cubed, or of mass per length per time squared,
or of inverse time squared.
Therefore, we may temporarily choose units of time so that $\beta=1$.
(This gives $\hbar \neq 1$, but that is irrelevant for our classical calculation.)
Then $V(\phi) = \phi$, $V' = 1$, and $V'' = 0$, so with $x=\phi_i-\phi$,
\begin{eqnarray}
{d^2K\over dx^2} = {1\over 4K}\left(1-{dK\over dx}\right)\left(1-3{dK\over dx}\right)
 + {9\over 2} K - 9\phi_i + 9x,
\label{eq:10}
\end{eqnarray}
with the boundary condition that $K = {1\over 3}x - {9\over 5}\phi_i x^2 + O(x^3)$
for $x \ll 1$.
Here $\phi_i$ is the single free parameter, which can be varied to get any desired
value for $w_0$ when $\Omega_m = \Omega_{m0}$.

When one evolves Eq. (\ref{eq:10}) numerically,
$K(x) \equiv {1\over 2}v^2 \equiv {1\over 2}\dot{\phi}^2 = {1\over 2}(\rho_s + P_s)$
increases monotonically with $x$, and $H \equiv \dot{a}/a = (1-dK/dx)/\sqrt{18K}$
decreases monotonically from $+\infty$ at the dust-dominated Big Bang to $-\infty$
at the stiff-scalar-dominated Big Crunch.

Take the past age to be
$t_0 = 10^8\mathrm{yr}/\alpha = 13.7 \mathrm{\ Gyr}$ and $\Omega_{m0} = 0.27$
\cite{WMAP}.
Then for a non-convex scalar field potential, the lower limit on the future lifetime
$t_c$ (until the Big Crunch) is a function of $w_0 = P_s/\rho_s
= (\mathrm{scalar\ field\ pressure})/(\mathrm{scalar\ field\ energy\ density})$,
with values given in the following table:\\

\begin{tabular}{lrrr}
$w_0$ & $R = {t_c\over t_0}$ & $\ \ \ \ \ {0.128\over 1+w_0}-1.68-3.81w_0$ &
 $\ \ \ \ t_c$ in Gyr \\
\hline
$-0.78$ & $1.90$ & $1.87$ & $26$ \\
$-0.86$ & $2.44$ & 2.51 & 33 \\
$-0.90$ & $2.99$ & 3.03 & 41 \\
$-0.95$ & $4.68$ & 4.50 & 64 \\
$-0.98$ & $9.21$ & 8.45 & 130 \\
$-0.99$ & $16.3\ \:$ & $14.9\ \:$ & 220 \\
$-0.999$ & $135\ \ \ \ $ & $130\ \ \ \ $ & $1\,800$ \\
$-0.9999$ & $1\,300\ \ \ \ $ & $1\,300\ \ \ \ $ & $18\,000$ \\
$-0.99999$ & $12\,900\ \ \ \ $ & $12\,800\ \ \ \ $ & $180\,000$ \\
$-0.999999$ & \ \ \ $128\,000\ \ \ \ $ & $128\,000\ \ \ \ $ & $1\,800\,000$ \\
\end{tabular}\\

The formula at the top of the third column is a crude fit to $R(\Omega_{m0},w_0)$
for $\Omega_{m0} = 0.27$; one can see from the table of values how accurate it is.

Therefore, if we assume that $V(\phi)$ is non-convex and take 
$t_0 = 13.7\ \mathrm{Gyr}$, $\Omega_{m0} = 0.27$, and $w_0 \leq -0.78$ \cite{WMAP},
we find a lower limit of the future lifetime of the universe of 26 Gyr,
or a minimum total lifetime of the universe of 40 Gyr.

If we had not only $w_0$ but also $(H\dot{w})_0$, the present value
of $H\dot{w}$, we could get lower limits on lifetimes with nonlinear potentials
$V(\phi)$ with some lower limits on, say, $V'''(\phi)$.

\section{Upper limits on the lifetime of the universe}

Can we also get a finite upper limit on the lifetime of the universe?
Apparently not from classical arguments and current observations,
which are consistent with a true cosmological constant $\Lambda > 0$,
$w_0 = -1$, and asymptotically de Sitter expansion forever.

But several theorists have argued that de Sitter space is actually unstable
\cite{Star,GKS,KKLT,Gid,FLW,GM,FS}.  The lifetime is argued to be
\begin{eqnarray}
\tau && < \mathrm{\ recurrence\ time} \ \sim \ (\mathrm{number\ of\ states})
 \ \sim \ e^S \ = \ e^{A/4} \ = \ e^{3\pi/\Lambda} \nonumber \\
 && = \exp{\left({\pi\over\Omega_{\Lambda}H^2}\right)}
     = \exp{\left({1.03\times 10^{122}\over\Omega_{\Lambda}h_0^2}\right)}
     = \exp{\left[2.80\times 10^{122}\left({0.73\over\Omega_{\Lambda_0}}\right)
            \left({0.71\over h_0}\right)^2\right]} \nonumber \\
 && \approx e^{e^{e^{e^{e^{.54823}}}}} \approx 10^{10^{10^{10^{.31945}}}}
  \gg\!\!\gg \mathrm{googleplex} = 10^{10^{10^2}}
  \approx e^{e^{e^{e^{e^{.52727}}}}} \approx 10^{10^{10^{10^{.30103}}}}.
\label{eq:11}
\end{eqnarray}

Here I shall develop a cosmological doomsday argument that we have observational
evidence for a lifetime $\ll\!\!\ll$ googleplex.

Assume an observation is described by a localized positive operator $\hat{A}$,
such as a projection operator onto some state of a brain with suitable environment
to permit the observation.  Then for a universe larger than the size of the
localized operator, $\langle\psi|\hat{A}|\psi\rangle > 0$ for virtually any state
$|\psi\rangle$, so the observation will have some positive probability of occurring
(e.g., as a vacuum fluctuation).  The only way out that I see would be if observers
and observations require the entire size of the universe, such as a projection
operator onto some total value of a globally conserved quantity.

Assume a human observer has finite size and just requires a 1 kg brain to last
0.1 second.  The dimensionless action for this is
\begin{eqnarray}
I(\mathrm{brain}) \sim {E\Delta t\over\hbar}
\sim {(1\: \mathrm{kg})(3\times 10^8\: \mathrm{m/s})^2(0.1\:\mathrm{s})
\over 10^{-34}\:\mathrm{J}\cdot\mathrm{s}} \sim 10^{50}.
\label{eq:12}
\end{eqnarray}
Therefore, in any spacetime volume
$V_4(\mathrm{brain}) \sim (10\:\mathrm{cm})^3 (0.1\:\mathrm{s})
\sim 10^{144}\: l^4_{\mathrm{Pl}} \sim e^{331}\: l^4_{\mathrm{Pl}}$,
one would expect a human observation to have probability
$P \geq P(\mathrm{brain}) \sim e^{-I} \sim e^{-10^{50}}$.
Thus observations should occur at a rate per Planck 4-volume of
\begin{eqnarray}
R \:>\: {P(\mathrm{brain})\over V_4(\mathrm{brain})}
 \sim e^{-331-10^{50}} \sim e^{-10^{50}}.
\label{eq:13}
\end{eqnarray}

Assuming
$N \sim (10^{11} \mathrm{\ people})(10 \mathrm{\ observations/s})(10^9 \mathrm{\:s})
\sim 10^{21} \sim e^{48}$ observations during past human history, we just need a
spacetime volume of
\begin{eqnarray}
V_4 \sim N R^{-1} < e^{48+331+10^{50}} \sim e^{10^{50}}
\label{eq:14}
\end{eqnarray}
to get more observations by vacuum fluctuations than have occurred during past human
history.

But if our spacetime lasts long enough to give a 4-volume (of our comoving part)
$V_4 \gg e^{I(\mathrm{brain})} \sim e^{10^{50}}$,
then almost all human observations would be from vacuum fluctuations.
However, these would almost entirely be much more disordered than what we experience.
Therefore, our ordered observations would be highly atypical in this scenario.

This extreme atypicality is like an extremely low likelihood, counting as very strong
observational evidence against any theory predicting such a long-lived universe
with a quantum state that can allow localized observations.

Thus I predict that the universe will not last long enough to give 4-volume
$> e^{10^{50}}$.

One gets various upper limits on the future lifetime from different assumptions about
the future expansion.  If all of the energy density of the universe, including the
dark energy, decays to radiation with $\rho \sim 1/a^4$, then a $k=0$ FRW universe
will asymptotically have $a \sim t^{1/2}$ and $V_4 \propto \int a^3 dt \sim t^{5/2}$,
which would exceed $e^{I(\mathrm{brain})} \sim e^{10^{50}}$ at
$t \sim \left(e^{10^{50}}\right)^{2/5} = e^{0.4\times 10^{50}} \sim e^{10^{49.6}}$.
If one has instead a $k=-1$ FRW universe, asymptotically $a \sim t$ and $V_4 \sim
t^4$, exceeding $e^{I(\mathrm{brain})}$ at
$t \sim \left(e^{10^{50}}\right)^{1/4} = e^{0.25\times 10^{50}} \sim e^{10^{49.4}}$.
Thus for any power-law expansion with exponent of order unity, I would predict that
the universe will not last past
\begin{eqnarray}
t \approx e^{10^{50}} \sim e^{e^{115}} \sim e^{e^{e^{e^{e^{.443}}}}}
    \ll\!\!\ll \mathrm{googleplex} \approx e^{e^{e^{e^{e^{.52727}}}}}.
\label{eq:15}
\end{eqnarray}

On the other hand, if the dark energy has a positive minimum energy density
($\Lambda > 0$), then the universe asymptotically has
$a \sim e^{H_\Lambda t} = e^{\sqrt{\Lambda/3} t}$, so
$V_4 \propto \int a^3 dt \sim \int e^{3H_\Lambda t} dt \sim e^{3H_\Lambda t}$,
which would exceed $e^{I(\mathrm{brain})} \sim e^{10^{50}}$ at
$t \sim I/(3H_\Lambda) \sim 10^{50}/(3H_\Lambda) \sim 10^{50}/\sqrt{3\Lambda}$.
For $H_\Lambda = \sqrt{\Lambda/3} = \sqrt{\Omega_{\Lambda_0}H_0^2}
= \sqrt{\Omega_{\Lambda_0}}h_0/(9.77813 \mathrm{\:Gyr}) \sim 1.06
\times 10^{-61}\: t_{\mathrm{Pl}}^{-1}$,
one gets $t \sim 10^{50}(5.4\times 10^9 \mathrm{\:yr}) \sim 10^{60} \mathrm{\:yr}
\sim 10^{111}\: t_{\mathrm{Pl}}$.

Therefore, if the current dark energy were due to a true cosmological constant,
it would produce too much spacetime volume to be consistent with our
non-vacuum-fluctuation (i.e., ordered) observations after only about $10^{60}$ years.

What are the implications for the string landscape or stringscape?
The observational evidence for $V_4 < e^{10^{50}}$ seems to imply that negligibly few
universes permitting observers would arise from positive metastable minima of the
stringscape potential with tunneling lifetime greater than about $10^{50}\:t_0 \sim
10^{111}\: t_{\mathrm{Pl}}$.

Is it possible that all life-permitting positive metastable minima decay away within
$10^{50}\:t_0 \sim e^{254}\: t_{\mathrm{Pl}}$?  Or, if longer-living ones exist, can
they possibly be overwhelmed by non-extreme regions that are nevertheless low and
flat enough for observers?  Or might it be true that the stringscape simply has
{\emph no} positive local minima at all?

For either of the latter two possibilities, we might speculate that by measurements of
$w(t) > -1$, we could observe our universe already sliding toward oblivion.

\section{CONCLUSIONS}

Thus I would predict, based on current observations and some plausible assumptions
(non-convex $V(\phi)$ and observers smaller than the universe),
that the future lifetime of our universe (at least in a form permitting observers)
has both lower and upper bounds:
\begin{eqnarray}
26\mathrm{\ Gyr}\ <\ t_{\mathrm{future}}\ <\ e^{10^{50}}\mathrm{\ Gyr\ or\ }
10^{51}\mathrm{\ Gyr},
\label{eq:16}
\end{eqnarray}
with the first possibility on the right being for future power-law expansion, and the
second being for exponential expansion at the rate given by the current dark energy.

Our observations also suggest that the dark energy of our universe is not near a
positive local minimum ($\Lambda > 0$), unless it can decay within about $10^{60}\:
\mathrm{yr} \sim 10^{111}\: t_{\mathrm{Pl}} \sim e^{254}\: t_{\mathrm{Pl}}$.
Furthermore, the string landscape or stringscape should not have \emph{any}
significant long-lived positive metastable minima with false vacua suitable for
observers (e.g., with lifetime longer than about $10^{60}\: \mathrm{yr}$ for a minimum
with roughly the same energy density as the observed dark energy), for such minima
would expand to give a huge volume and too many observers from vacuum fluctuations to
be consistent with our ordered observations being typical.

\begin{acknowledgments}

I greatly appreciated the hospitality of the George P. and Cynthia W. Mitchell
Institute for Fundamental Physics of the Texas A \& M University, where most of this
work was prepared and first presented on 2004 March 17.  I am also thankful for the
hospitality of Kunsan National University in Gunsan, South Korea, the Asia Pacific
Center for Theoretical Physics in Pohang and Seoul, South Korea, and the 9th
Italian-Korean Symposium on Relativistic Astrophysics in Seoul, South Korea, and Mt.
Kumgang, North Korea, where the present lecture was given on 2005 July 22 to an
audience that unfortunately could not include any North Koreans.

I am grateful for conversations with Cliff Burgess, George Efstathiou, Willy
Fischler, Valeri Frolov, Gary Gibbons, Alan Guth, Jim Hartle, Thomas Hertog,
Gary Horowitz, Shamit Kachru, Renata Kallosh, Sang Pyo Kim, Andrei Linde, Hong
L\"{u}, Burt Ovrut, Joe Polchinski, Christopher Pope, Eva Silverstein, Alexei
Starobinsky, Andrew Strominger, Leonard Susskind, Bill Unruh, Bob Wald, and
many others who may or may not have had an influence on this paper, though I
would not want to blame anyone other than myself for the second part.

This work was partially supported by Korea Astronomy and Space Science Institute
through Kunsan National University, and by the Natural Sciences and Engineering
Research Council of Canada.

\end{acknowledgments}



\end{document}